\documentclass[twocolumn,showpacs,preprintnumbers,prl,amsmath,amssymb]{revtex4}
\usepackage{graphicx}
\usepackage{dcolumn}
\usepackage{bm}

\begin{document}



\title{Transport in Ultraclean YBa$_2$Cu$_3$O$_7$: 
neither Unitary nor Born Impurity Scattering }

\author{R.W.Hill$^1$, Christian Lupien$^{1\star}$, M.Sutherland$^1$, Etienne 
Boaknin$^1$, D.G.Hawthorn$^1$, Cyril Proust$^{1\dagger}$, \\
F.Ronning$^1$, Louis Taillefer$^{1,2}$, 
Ruixing Liang$^{3}$, D.A.Bonn$^{3}$ and W.N.Hardy$^{3}$}

\affiliation{$^1$Department of Physics, University of Toronto, Toronto,
Ontario, Canada\\
$^2$D\'{e}partement de physique, Universit\'{e} de
Sherbrooke, Sherbrooke, Qu\'{e}bec, Canada\\
$^3$Department of Physics, University of British Columbia, BC,
Canada}


\date{July 15 2003}

\begin{abstract}

The thermal conductivity of ultraclean YBa$_2$Cu$_3$O$_7$ 
was measured at very low temperature in magnetic fields up to 
13 T.  
The temperature and field dependence of the electronic heat conductivity
show that two widespread assumptions of transport theory applied to
unconventional superconductors fail for clean cuprates:
impurity scattering cannot be treated in the usual unitary limit (nor
indeed in the Born limit),
and scattering of quasiparticles off vortices cannot be neglected.
Our study also sheds light on the long-standing puzzle of a sudden onset of
a "plateau"  in the thermal conductivity of Bi-2212 versus field.

\end{abstract}


\maketitle


Since 1985, the theory of transport in unconventional superconductors 
has been dominated by the ubiquitous assumption that impurity scattering 
must be treated in the unitary limit ($\pi/2$ phase shift) \cite{Pethick}. 
The aim of this Letter is to test this assumption in the simplest and most
reliable context available, namely that of the cuprate material
YBa$_2$Cu$_3$O$_x$ (YBCO), for which the superconducting gap structure is
firmly established and straightforward, with $d_{x^2-y^2}$ symmetry over an
approximately cylindrical Fermi surface.  We do this by measuring heat
transport in ultraclean single crystals that have a scattering rate one
order of magnitude lower than in previous studies. This allows us to
reliably resolve the intrinsic temperature dependence of the electronic
thermal conductivity.

The pivotal result of transport theory is the {\em universal limit of
conductivity}, whereby the ability of a superconductor to carry heat as
$T\to0$ is independent of either the concentration of scattering centers or
the strength of the scattering potential ({\it i.e.} phase shift).
The universal limit was first observed in YBCO \cite{Taillefer2}, and then
confirmed in Bi$_2$Sr$_2$CaCu$_2$O$_8$ (Bi-2212) \cite{Behnia}, as well as
in the $p$-wave superconductor Sr$_2$RuO$_4$ \cite{Suzuki}.
It is only by going beyond the universal limit that one can test the
assumption of unitary scattering.  The increase in conductivity with
temperature $T$ or applied magnetic field $H$ is expected to depend
strongly on scattering phase shift $\delta$ \cite{Graf,Kubert}.  The main 
finding of the present study is that the
electronic thermal conductivity increases
with $T$ much faster than could ever be expected from unitary scattering.
The high degree of order in the present crystal also reveals unambiguously
the dominance of vortex scattering over impurity scattering even at modest
fields.  This will shed light on the long-standing puzzle of a sudden onset
of a ``plateau'' observed previously in the thermal conductivity of Bi-2212
versus field \cite{Krishana, Aubin}.



The thermal conductivity $\kappa$ was measured using a single heater-two
thermometer method.  The heat current was supplied along the $a$-axis of
the sample, and the magnetic field applied parallel to the $c$-axis.  The
measurements were made in a dilution refrigerator by varying the
temperature from 0.04~K to $> 0.7$~K at fixed magnetic field.  The samples
were field-cooled by cycling to $T>100$~K before changing the field.  The
error in the absolute value of the conductivity is estimated to be
approximately $10\%$.  The relative error between temperature sweeps at
different fields is of order 1\%.


The single crystal platelet of YBCO used in this study has dimensions
$1.0\times0.5$ mm$^2$ and 25 $\mu$m thick.  It was grown in a BaZrO$_3$
(BZO) crucible \cite{Liang}, which results in crystals with extremely
high chemical purity (99.99 - 99.995\%) and a high degree of
crystalline perfection as compared with crystals grown in Y$_2$O$_3$-stabilised 
ZrO$_2$ (YSZ) crucibles. The sample was detwinned at
250$^\circ$C under uniaxial stress, followed by a 50 day annealing 
process at 350$^\circ$C, resulting in oxygen chains with less than 0.7\%
vacancies.  This level of oxygen doping, $x = 6.99$, is slightly
above that for maximal $T_c$ (93 K), resulting in a marginally lower
$T_c = 89$ K.  Four electrical contacts were applied using silver
epoxy fired at 350$^\circ$C for 1 hour, 
giving typical contact resistances of 200~m$\Omega$ at 4.2~K.
  
\begin{figure}
\centering
\resizebox{\columnwidth}{!}{
\includegraphics{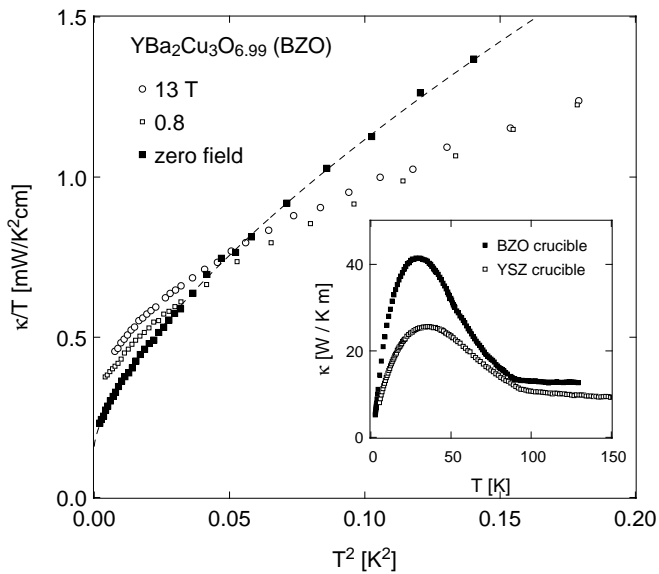}}
\caption{\label{fig:KlowT} Thermal conductivity divided by
temperature versus $T^2$ for a high-purity YBCO sample for magnetic
fields applied perpendicular to the $ab$ plane.  The dashed line is a 
guide to the eye.
Inset: High-temperature data for 
YBCO samples grown in different crucibles.}
\end{figure}

The high purity of our sample can be established from its thermal 
conductivity at high temperature.  A low 
defect level leads to a larger peak in 
$\kappa$ below $T_c$, as noted by Zhang {\it et al} \cite{ZhangY}.  In the
inset to Fig.~\ref{fig:KlowT}, we show the thermal conductivity for
this sample as compared to a detwinned YBCO sample 
with $x~=~6.95$ grown in an YSZ crucible.  The approximate doubling in
peak height can be attributed to an order of magnitude decrease in
the intrinsic disorder level \cite{Hirshfeld5} (see also \cite{Sutherland}).

\begin{figure}
\centering
\resizebox{\columnwidth}{!}{
\includegraphics{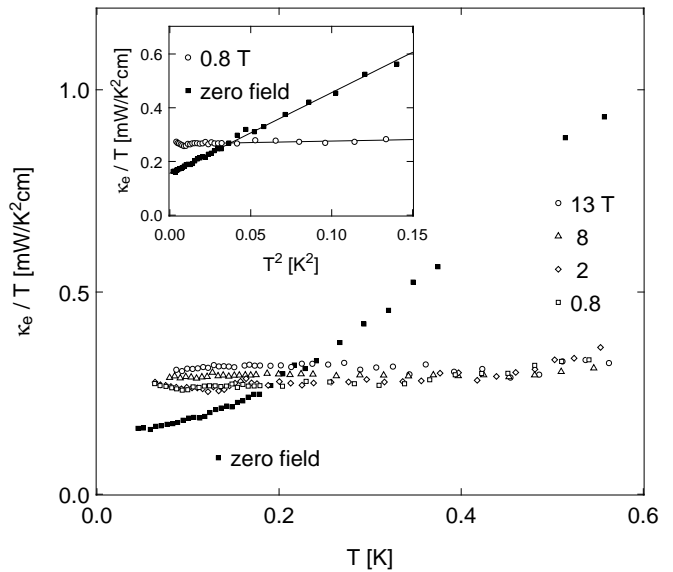}}
\caption{\label{fig:K_e} Electronic thermal conductivity, plotted 
as $\kappa_e/T$ vs $T$, for magnetic fields from zero to 13 T.  
The electronic contribution is
extracted as discussed in the main text. Inset: The low field curves 
plotted against $T^2$.}
\end{figure}

The main panel of Fig.~\ref{fig:KlowT} shows the temperature
dependence of the thermal conductivity of the high-purity (BZO)
sample at zero field and in applied fields of 0.8 and 13 T.  
The thermal conductivity is the sum of electronic and phononic 
contributions: $\kappa = \kappa_e + \kappa_{ph}$.  The residual linear 
term, $\kappa_0/T$, obtained by extrapolating $\kappa/T$ to 
$T\rightarrow0$, is entirely electronic.  The question is how to 
extract any $T$-dependence of $\kappa_e/T$.  We use a magnetic field 
to do this. 
First, note that the zero-field curve in Fig.~\ref{fig:KlowT} shows
a more rapid increase with temperature than the two in-field curves
which are approximately parallel.
On the assumption that the phonon transport at very low temperatures
is limited by scattering from the boundaries of the sample (see 
\cite{Sutherland}) and is
unaffected by scattering from vortices, we can only attribute this 
difference to 
electrons.  Furthermore, since all subsequent in-field curves lie 
parallel we assume that this additional electronic conduction is 
completely 
suppressed when a magnetic field is applied (see inset of
Fig.~\ref{fig:K_e}), and the remaining 
temperature dependence of $\kappa/T$ is due entirely to phonons.   
In other words $\kappa_{ph}$ is the $T$-dependent part of the 13 T 
data: 
$\kappa_{ph}/T = \kappa(13$ T$)/T - \kappa_0(13$ T$)/T$, 
where $\kappa_0(13$ T$)/T = 0.31$ mW/K$^2$cm.  
The electronic conductivity,
$\kappa_e/T$, is then given by subtracting this from the total conductivity: 
$\kappa_e(H,T)/T = \kappa(H,T)/T - 
\kappa_{ph}/T$. This is shown in Fig.~\ref{fig:K_e} for 
applied magnetic fields from 0-13 T.  

The zero-field electronic conductivity shows a rapid 
growth with temperature, increasing by a factor of five within 0.5 K.  
The inset plots the same data on a $T^2$ temperature scale showing 
that the growth is cubic in temperature.  
%
As soon as a magnetic field 
is applied this temperature dependence is completely suppressed (see inset 
of Fig.~~\ref{fig:K_e}), an 
effect which can only be attributed to the scattering of 
quasiparticles by vortices.  
%
The possibility that the change in temperature dependence between
zero and applied field is due to scattering of phonons by vortices is
ruled out by the lack of field dependence above 0.8 T.  By
increasing the field to 13 T, an order of magnitude more
vortices have been introduced to the system, yet the total 
conductivity remains essentially unchanged.  
%


{\em Zero magnetic field ($H$=0)}.
Using a self-consistent quasiclassical theory, formulated at low
temperatures where heat transport is limited by electron scattering from
random defects \cite{Hirshfeld4, Hirshfeld3}, the thermal conductivity due
to quasiparticles at the nodes of a $d$-wave superconductor is given by
\cite{Graf}
\begin{equation}
\frac{\kappa_e}{T}\left(T\right) = \frac{\kappa_{00}}{T}
\left[1 + \frac{7\pi^2}{15}\left(\frac{a^2T}{\gamma}\right)^2\right]
\label{eq:gamma}
\end{equation}
\noindent where $\kappa_{00}/T$ is the universal conductivity limit,
$\gamma$ is the impurity bandwidth and the coefficient $a$ is strongly
dependent on the scattering phase shift.  This expression is valid in the
dirty limit where $k_BT<\gamma$. The temperature dependence of our
extracted electronic conductivity is well described by this form.  Fitting
this expression to the zero field data (see inset of Fig.~\ref{fig:K_e})
gives $\kappa_e/T = 0.16 (1 + 19.2 T^2)$.  Note that this corresponds to a
huge 20-fold increase in $\kappa_e/T$ by 1~K.

{\em $H=0$: Zero temperature}.
The residual linear term, $\kappa_{0}/T =
0.16$~mW/K$^2$cm, is in excellent agreement with the value published
for optimally doped YBCO ($0.14$~mW/K$^2$cm~\cite{Taillefer2}).  
{\em This is direct confirmation of the universal nature of low
temperature quasiparticle thermal conductivity.} In this case we have
measured a sample in the more difficult regime of increased purity,
where we observe an order 
of magnitude lower scattering rate
than previously, and still recover the universal limit.
(A slight increase in $\kappa_0/T$ is expected based on the small increase
in doping from $x=6.95$ to $x=6.99$ \cite{Sutherland}.)

{\em $H=0$: Finite temperature}.
From the temperature dependent part of our fitted data we can 
estimate the impurity bandwidth and scattering rate.
For scattering in the unitary limit $a = 1/2$ and the impurity bandwidth
$\gamma$ is related to the normal-state scattering rate $\Gamma_n$ by the
relation $\gamma = 0.63 \sqrt{\Delta_0\Gamma_n}$ \cite{Hirshfeld3}.  Using the 
value from the fitted zero-field data (inset of Fig.~\ref{fig:K_e}),
we obtain
$\gamma\sim0.25$~K and 
using $\Delta_0 = 2.14~k_B T_c$, we get
$\Gamma_n/T_c\sim10^{-5}$, therefore
$\Gamma_n\sim1\times10^{8}$~s$^{-1}$.  
Such a small scattering rate is unrealistic.
Using $v_F = 2.5\times10^7$~cm/s \cite{Sutherland}, it would imply a
normal-state mean free path as long as the longest dimension of the sample:
$l\sim1$~mm!

In the Born approximation, $a = (\pi v_2 \tau_0)/2$ and $\gamma =
4\Delta_0\exp(-\pi \Delta_0/2\Gamma_n)$.  Assuming a pure $d$-wave gap
gives $v_2 = 2\Delta_0/\hbar k_F$ and using $\tau_0 = 1/2\Gamma_n$, leads 
to
$\gamma\sim3$ K and 
$\Gamma_n\sim0.6 \Delta_0\sim2.5\times10^{12}$~s$^{-1}$.  
Again we estimate a
scattering rate that is unrealistic, 
in this case much too large.
If the scattering rate were truly this magnitude it would lead to a
substantial suppression of $T_c$, as noted previously \cite{Hirshfeld3},
which is not observed experimentally.

{\em From the measured temperature dependence of the electronic thermal
conductivity, we conclude that the usual quasiclassical calculation cannot
be correct if it treats impurity scattering with a single isotropic phase
shift of either 0 (Born) or $\pi/2$ (unitary).}


In a broader context, similar measurements on unconventional
superconductors UPt$_3$ \cite{Suderow} and Sr$_2$RuO$_4$ \cite{Suzuki}
reveal conductivities far too small to consider Born scattering.  They are
therefore analysed in the unitary limit.  Consistent with the present work,
the trend that emerges is that a quantitative analysis of transport in the
superconducting state, using the unitary limit, leads to mean free paths
considerably longer than are consistent with normal state measurements such
as de Haas van Alphen studies or resistivity above $H_{c2}$.  This suggests
that the standard theoretical approach to transport in unconventional
superconductors is generically inadequate.

Microwave conductivity measurements \cite{Hosseini, Turner} on samples of
identical quality to that used in this study have also been compared to
current theories of transport in unconventional superconductors
\cite{Berlinski}.  These measurements reflect {\it some} but not all of the
characteristics of weak scattering (Born limit) and point to either
inadequacies in the conventional theories, or the need to consider
intermediate phase shifts, or both.

An earlier study on Bi-2212 reported an electronic conductivity,
$\kappa_e/T$, that increased as $T$ rather than $T^2$ \cite{Behnia}.  The
phonon contribution is subtracted by comparing measurements before and
after electron irradiation.  A linear $T$-dependence is expected in the
clean limit, where $\gamma < k_B T$.  It is unclear, however, why Bi-2212
crystals with a level of disorder two orders of magnitude larger than our
BZO-grown YBCO crystals should be in that limit.  Indeed it is a surprise
that any $T$-dependence of $\kappa_e/T$ can be resolved in such crystals.  
(A 2-fold increase is observed by 1~K).

\begin{figure}
\centering
\resizebox{\columnwidth}{!}{
\includegraphics{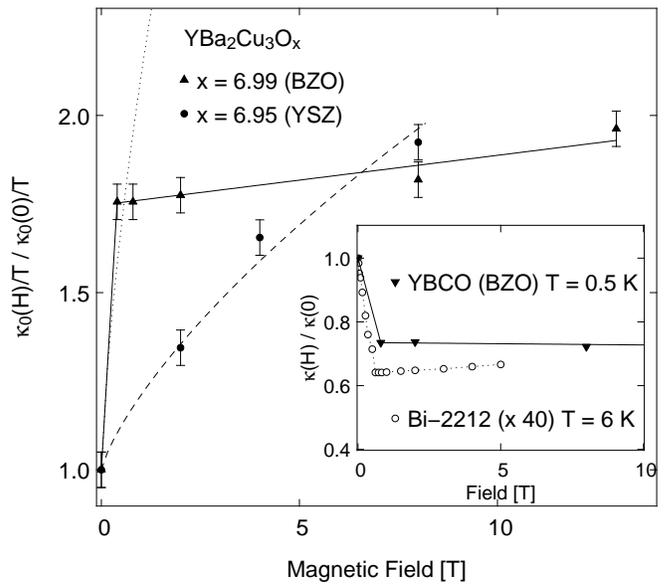}}
\caption{\label{fig:KvsH} Normalized residual electronic conductivity as a 
function of magnetic field for a slightly overdoped ($x=6.99$), very 
pure, detwinned sample of YBCO (triangles).  
For comparison, the same is plotted for optimally-doped YBCO ($x=6.95$),
where the sample is less pure \cite{Chiao}, along with the fit to a
semiclassical theory which describes the data \cite{Kubert} (dashed line).  
An attempt to fit the same theory to the BZO-grown sample is also shown
(dotted line).
Inset: 
High temperature isotherm at 0.5 K for the pure sample compared to a 
magnified ($\times$ 40)
plot of an isotherm at 6 K for Bi-2212 \cite{Krishana}.}
\end{figure}

{\em Finite magnetic field ($H>0$)}.
In Fig.~\ref{fig:KvsH} the extrapolated linear thermal 
conductivity at 
$T\rightarrow0$ is plotted as a funtion of magnetic field.  The values 
are normalised by the zero-field value $\kappa_{0}/T$.
Also plotted is the previous data for an optimally-doped YBCO ($x=6.95$)
sample grown in a YSZ crucible \cite{Chiao}.  In contrast to the latter
sample, the much purer one shows a rapid increase at fields below $H = 0.4$
T followed by a sudden change to a regime where its behaviour is almost
field independent.  Concomittantly, the coefficient of the $T^2$ term, and
consequently the quasiparticle lifetime, collapses (see inset of
Fig.~\ref{fig:K_e}). We interpret this as an indication that
quasiparticle-vortex scattering is strongly influencing the transport in
this very clean material.

Current semi-classical treatments of the effect of magnetic field on the
low temperature electronic thermal conductivity predict smooth sublinear
increases to within logarithmic corrections \cite{Kubert, Vekhter}. This
arises when quasiparticle energies are doppler shifted by the
superfluid flow around magnetic vortices.  Earlier measurements, reproduced
in Fig.~\ref{fig:KvsH}, on YSZ-grown samples \cite{Chiao} are well
described by such theories and have been viewed as additional evidence for
$d$-wave symmetry of the superconducting gap. Clearly the field dependence
of the measurements reported here on a BZO-grown sample is of completely
different character and cannot be described by available theories.  To
emphasize this the dotted line in Fig.~\ref{fig:KvsH} is the best fit for
the theory of K\"{u}bert and Hirschfeld \cite{Kubert} to the initial rise
of the conductivity.  Intriguingly the scattering rate deduced from this
fit is an order of magnitude smaller than for the YSZ sample in agreement
with the relative change inferred from the rise in thermal conductivity
below $T_c$.  This suggests that the rapid growth at low-fields, due to the
low impurity scattering rate, is truncated by an additional scattering
mechanism excluded from current theories.

A common feature of these theories is that at low temperatures the effect
of scattering from vortices is expected to be negligible in comparison to
impurity scattering.  However given the demonstrably low impurity
scattering rate in this BZO-grown sample we argue that such a
simplification may be incorrect. Assuming these scattering mechanisms to be
independent, one could add them in a Matthiessen-like manner.  At zero
field the conductivity will be in the impurity dominated regime, whilst in
the very high field limit, vortices will dominate the scattering.  The
crossover between these two limits will depend on the relative amount of
impurity scattering and the cross-section for vortex scattering.  Such a
model may also account for the similar magnitude of the conductivity at
high fields for the two samples shown in Fig.~\ref{fig:KvsH}, despite their
obviously different impurity concentrations.

A similar strategy was adopted by Franz \cite{Franz} to include
scattering from a random vortex lattice.  In this case the theory was
developed to explain a plateau-ing of the field dependence of the
thermal conductivity at higher temperatures and has yet to be
extended to low enough temperatures for comparison with the data
here.  Nevertheless, the phenomenology may be correct; a
$\sqrt{H}$ increase coming from a doppler shift is exactly
compensated by scattering from vortices, where the scattering length
is given by the inter-vortex separation $\sim\sqrt{H}$.  What remains
unusual is the sharpness with which this compensation turns on.

With this is mind, the inset to Fig.~\ref{fig:KvsH} shows the field
dependence of an isotherm at 0.5 K. The behaviour is reminiscent of earlier
measurements on Bi-2212 \cite{Krishana}, also shown as an isotherm at 6 K,
where the high-field ``plateau'' is preceded by a ``kink''.  Such
similarity, measured here for the first time in YBCO, is suggestive of a
common origin.  Since the present work is on a sample of the highest purity
and in field-cooled measurements, this phenomena cannot simply be dismissed
as material dependendent extrinsic behaviour, nor solely related to gap
anisotropy \cite{Ando}.  The hysteresis seen in Bi-2212 measurements 
\cite{Aubin} is
also naturally explained by the dominance of vortex scattering. The
dramatic difference in magnitude between these two measurements is likely a
consequence of the small electronic conductivity relative to the phonon
contribution (which has not been subtracted) at 6 K in Bi-2212 and the
comparatively huge electronic contribution measured in this high-purity
YBCO.

An alternative explanation has recently been offered by Franz and Vafek
\cite{Franz3}. In their fully quantum-mechanical theory, the Meissner state
(at zero field) and the vortex state emerge as two distinct $d$-wave states
with different quasiparticle effective velocities. They both exhibit
universal conductivity, with different values of the universal limit (see
Fig.~\ref{fig:K_e}). This appealingly accounts for the fact that the
conductivity of the two different YBCO samples is the same not only at zero
field but also at high fields (see Fig.~\ref{fig:KvsH}). It is not clear,
however, why the finite temperature correction to this universal limit
should be so dramatically different in the two states.


In conclusion, when analyzed in terms of the quasiclassical theory of
transport for a $d$-wave superconductor, the thermal conductivity of an
extremely pure sample of YBCO reveals two features:
1) the universal limit as $T \to 0$ is confirmed, 2) the usual assumption
that impurity scattering can be treated as single isotropic phase shift in
the unitary limit (or the Born limit) is incorrect. Transport theory as it
stands must be revised, at least in the clean limit, perhaps by going to
intermediate phase shifts and maybe in more profound ways.
Moreover, in the presence of a magnetic field, we find that transport
appears to be rapidly dominated by vortex scattering, which can therefore
not be neglected as it usually is.

\begin{acknowledgments}

We would like to thank R.Gagnon for help with the samples, and 
J.Carbotte, A.Durst, M.Franz,  R.Harris, P.Hirschfeld, F.Marsiglio and 
A.Vishwanath
for stimulating
discussions. This work was supported by the Canadian
Institute for Advanced Research and funded by NSERC of Canada.\\
Present addresses: $^{\star}$LASSP, Cornell University, Ithaca, NY 14853, USA, 
$^{\dagger}$Laboratoire National des Champs Magn\'etiques Puls\'es,
143 avenue de Rangueil, 31432 Toulouse, France.

\end{acknowledgments}

\bibliography{YBCO_H_shorter}

\end{document}